\documentclass{article}
\usepackage[english]{babel} 

\usepackage{graphicx}

\usepackage{color}

\begin{document}

\newcommand{\rum}{\rule{0.5pt}{0pt}}
\newcommand{\rub}{\rule{1pt}{0pt}}
\newcommand{\rim}{\rule{0.3pt}{0pt}}
\newcommand{\numtimes}{\mbox{\raisebox{1.5pt}{${\scriptscriptstyle \rum\times}$}}}
\newcommand{\numtimess}{\mbox{\raisebox{1.0pt}{${\scriptscriptstyle \rum\times}$}}}
\newcommand{\Boldsq}{\vbox{\hrule height 0.7pt
\hbox{\vrule width 0.7pt \phantom{\footnotesize T}%
\vrule width 0.7pt}\hrule height 0.7pt}}
\newcommand{\two}{$\raise.5ex\hbox{$\scriptstyle 1$}\kern-.1em/
\kern-.15em\lower.25ex\hbox{$\scriptstyle 2$}$}

\renewcommand{\refname}{References}
\renewcommand{\tablename}{\small Table}
\renewcommand{\figurename}{\small Fig.}
\renewcommand{\contentsname}{Contents}

\begin{center}
{\Large\bf 
 Experimental Investigation of the Fresnel Drag Effect in RF Coaxial Cables\rule{0pt}{13pt}}\par

\bigskip
Reginald T. Cahill \\ 
{\small\it  School of Chemical and Physical  Sciences, Flinders University,
Adelaide 5001, Australia\rule{0pt}{15pt}}\\
\raisebox{+1pt}{\footnotesize E-mail: Reg.Cahill@flinders.edu.au}\par

\bigskip

David  Brotherton \\
{\small\it  Geola Technologies Ltd, Sussex Innovation Centre, Science Park Square, Falmer, East Sussex, BN1 9SB, United Kingdom\rule{0pt}{15pt}}\\
\raisebox{+1pt}{\footnotesize E-mail:  dbr@geola.co.uk}\par

\bigskip

{\small\parbox{11cm}{%
An experiment that confirms the Fresnel drag formalism in RF coaxial cables is reported. The Fresnel `drag' in bulk dielectrics and in optical fibers has previously been well established.  An explanation for this formalism is given, and it is shown that there is no actual drag phenomenon, rather  that the Fresnel drag effect is merely the consequence of  a simplified description of EM scattering within a dielectric in motion wrt the dynamical 3-space.  The Fresnel drag effect plays a critical role in the design of  various light-speed anisotropy detectors.    \rule[0pt]{0pt}{0pt}}}\medskip
\end{center}

\setcounter{section}{0}
\setcounter{equation}{0}
\setcounter{figure}{0}
\setcounter{table}{0}

\markboth{Cahill and Brotherton.  Experimental Investigation of the Fresnel Drag Effect in RF Coaxial Cables}{\thepage}
\markright{Cahill and and  Experimental Investigation of the Fresnel Drag Effect in RF Coaxial Cables }


\section{Introduction}

In 2002 it was discovered that the Michelson-Morley 1887 light-speed anisotropy  experiment \cite{MM}, using the interferometer in gas mode,   had indeed detected anisotropy, by taking account of both a physical  Lorentz length contraction effect  for the interferometer arms, and the refractive index effect of the air in the light paths  \cite{MMCK,MMC}.  The observed fringe shifts corresponded to an anisotropy speed in excess of 300km/s.  While confirmed by numerous later experiments, particularly that of Miller  \cite{Miller},  see \cite{CahillNASA} for an overview, the most accurate analysis used the Doppler shifts from spacecraft earth-flybys  \cite{And2008, CahillNASA}, which gave the solar-system a galactic  average speed through 3-space of   486km/s in the direction RA=4.29$^h$, Dec=-75.0$^{\circ}$, a direction within 5$^{\circ}$ of that found by Miller in his 1925/26 gas-mode Michelson interferometer experiment\footnote{This speed and direction is very different to the CMB speed and direction - which is an unrelated phenomenon.}.  In vacuum mode a Michelson interferometer cannot detect the anisotropy, nor its turbulence effects, as shown by the experiments in \cite{cavities}, actually using resonant orthogonal cavities.  These experiments show, overall, the difference between Lorentzian Relativity (LR)  and Special Relativity (SR).  In LR the length contraction effect is caused by motion of a rod, say,  through the dynamical 3-space, whereas in SR the length contraction is only a perspective effect,  occurring only when the rod is moving relative to an observer.  This was further clarified  when an exact mapping between Galilean space   and time coordinates and the Minkowski-Einstein spacetime coordinates was recently discovered \cite{CahillMink}.  This demonstrates that the SR time dilation and space contraction  effects are merely the result of using an unphysical choice of space and time coordinates that, by construction, makes the speed of light in vacuum an invariant, but only wrt to that choice of coordinates. Such a contrived  invariance  has no connection with whether light speed anisotropy is detectable or not - that is to be determined by experiments.  

The detection of light speed anisotropy - revealing a flow of space past the detector,  is now entering an era of precision   measurements. These are  particularly important because experiments have shown large turbulence effects in the flow, and  are beginning to characterise this turbulence. Such turbulence can be shown to correspond to what are, conventionally, known  as gravitational waves, although not those implied by General Relativity, as they are much larger than these \cite{Book, Review,CahillGW2009}.

The detection and characterisation of these wave/ turbulence effects requires the development  of small and cheap detectors, such as optical fiber Michelson interferometers \cite{CahillOF}. However in all detectors the EM signals travel through a dielectric, either in bulk or optical fiber or through RF coaxial cables. For this reason it is important to understand the so-called Fresnel drag effect.  In optical fibers the Fresnel drag effect has been established \cite{Wang}. This is important in the operation of  Sagnac optical fiber gyroscopes, for then the calibration is independent of the fiber refractive index. The Fresnel drag speed  is a phenomenological formalism that characterises the effect of the absolute motion of the propagation medium  upon the speed of the EM radiation within that medium.

The Fresnel drag expression is that a dielectric in absolute motion through space at speed $v$ causes the EM radiation to travel at speed
\begin{equation}
V(v)=\frac{c}{n}+v\left(1-\frac{1}{n^2}\right)
\label{eqn:Fresnel}\end{equation}
wrt the dielectric, when $V$ and $v$ have the same  direction. Here $n$ is the dielectric refractive index. The 2nd term is known as the Fresnel drag, appearing to show that the moving dielectric ``drags" the EM radiation, although below we show that this is a misleading interpretation. That something unusual was happening followed from the discovery of  stellar aberration by Bradley in 1725. Here the direction of the telescope must be varied over a year when observing a given star.   This is caused by the earth's orbital speed of  30km/s.  Then  Airy in 1871 demonstrated that the same aberration angle occurs even when the telescope is filled with water.  This effect is explained by the Fresnel expression in (\ref{eqn:Fresnel}), which was also confirmed by the Fizeau experiment  in 1851, who used two beams of light travelling through  two tubes filled with flowing water, with one beam flowing in the direction of the water, and the other counterflowing. Interferometric  means permitted the measurement of the travel time difference between the two beams, confirming  (\ref{eqn:Fresnel}), with $v$ the speed of the water flow relative to the apparatus.  This arrangement cannot detect the absolute motion of the solar system, as this contribution to the travel time difference cancels because of the geometry of the apparatus.

There have been various spurious ``derivations" of  (\ref{eqn:Fresnel}), some attempting to construct a physical ``drag" mechanism, while another uses the SR addition formula for speeds. However that well-known addition formula is merely a mathematical manifestation of using the unphysical Minkowski-Einstein coordinates noted above, and so is nothing but a coordinate effect, unrelated to  experiment. Below we give a simple heuristic derivation which shows that there is no actual ``drag" phenomenon.  But first we show the unusual consequences of   (\ref{eqn:Fresnel}) in one-way speed of EM radiation experiments. It also plays a role in 2nd order $v/c$ experiments, such as the optical-fiber Michelson interferometer \cite{CahillOF}.

 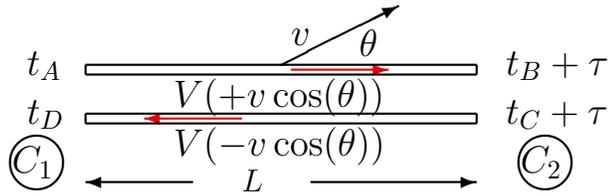
\begin{figure}
\vspace{20mm}
\hspace{30mm}
\setlength{\unitlength}{1.3mm}
\hspace{0mm}\begin{picture}(0,0)
\thicklines

\definecolor{hellgrau}{gray}{.8}
\definecolor{dunkelblau}{rgb}{0, 0, .9}
\definecolor{roetlich}{rgb}{1, .7, .7}
\definecolor{dunkelmagenta}{rgb}{.9, 0, .0}
\definecolor{green}{rgb}{0, 1,0.4}
\definecolor{black}{rgb}{0, 0, 0}

\color{dunkelmagenta}
\put(30,5.5){\vector(1,0){10}}
\put(25,0.5){\vector(-1,0){10}}

\color{black}

\put(9,5){\line(1,0){40}}
\put(9,6){\line(1,0){40}}

\put(9,5){\line(0,1){1}}
\put(49,5){\line(0,1){1}}

\put(9,0){\line(1,0){40}}
\put(9,1){\line(1,0){40}}

\put(9,0){\line(0,1){1}}
\put(49,0){\line(0,1){1}}

\put(3,5){\Large{$t_A$}}
\put(3,0){\Large{$t_D$}}

\put(52,5){\Large{$t_B+\tau$}}
\put(52,0){\Large{$t_C+\tau$}}

\put(29,6){\vector(2,1){12}}
\put(30,8){\Large{$v$}}
\put(37,6.9){\Large{$\theta$}}

\put(18,1.7){\Large{$V(+v\cos(\theta))$}}
\put(18,-3.3){\Large{$V(-v\cos(\theta))$}}

\put(4,-4){\circle{5}}
\put(1.5,-5){\Large{$C_1$}}

\put(56,-4){\circle{5}}
\put(53.5,-5){\Large{$C_2$}}

\put(20,-6){\vector(-1,0){11}}
\put(25,-7){\large$ L$}
\put(32,-6){\vector(1,0){17}}

\end{picture}

\vspace{12mm}
	\caption{\small{ Schematic layout for measuring the one-way speed of light in either free-space, optical fibers or RF coaxial cables, without requiring the synchronisation of the clocks $C_1$ and $C_2$.  Here $\tau$ is the unknown offset time between the clocks, and  $t_{A}, t_{B}+\tau,  t_{C}+\tau,   t_{D}$  are the observed clock  times, while  $ t_{B},  t_{C}$ are, {\it a priori}, unknown true times. $V$ is the light speed in (\ref{eqn:Fresnel}), and $v$ is the speed of the apparatus through space, in direction $\theta$.}}
 \label{fig:oneway}
\end{figure}

\section{One-way Speed of Light Anisotropy Measurements}
Fig.\ref{fig:oneway} shows the arrangement for measuring the one-way speed of light, either in vacuum, a dielectric, or RF coaxial cable.  It is usually argued that one-way speed of light measurements are not possible because the clocks cannot be synchronised.   Here we show that this is false, and at the same time show an important  consequence of  
(\ref{eqn:Fresnel}). In the upper part of  Fig.\ref{fig:oneway} the actual travel time $t_{AB}$ from $A$ to $B$ is determined by
\begin{equation}
V(v\cos(\theta))t_{AB}=|{\bf L}^\prime|
\end{equation}
where $|{\bf L}^\prime | = |{\bf L}+{\bf v}t_{AB}|\approx L+v\cos(\theta)t_{AB}+..$ is the  actual distance travelled,   at speed $V(v\cos(\theta))$, using $v t _{AB} \ll  L$, giving 
\begin{equation}
V(v\cos(\theta))t_{AB}=L+v\cos(\theta)t_{AB}+...
\label{eqn:traveltime1}\end{equation}
where the 2nd term comes from, approximately,  the end $B$ moving an additional distance  $v\cos(\theta)t_{AB}$ during the true time interval $t_{AB}$.  This gives
\begin{equation}
t_{AB}\approx \frac{L}{V(v\cos(\theta))-v\cos(\theta)}=\frac{nL}{c}+\frac{v\cos(\theta)L}{c^2}+..
\label{eqn:traveltime2}\end{equation}
on using (\ref{eqn:Fresnel}) and expanding to 1st oder in $v/c$.  
If we ignore the Fresnel drag term in (\ref{eqn:Fresnel}) we obtain, instead,
\begin{equation}
t_{AB}\approx \frac{L}{c/n-v\cos(\theta)}=\frac{nL}{c}+\frac{n^2v\cos(\theta)L}{c^2}+..
\label{eqn:traveltime3}\end{equation}
 The 1st important observation is that the $v/c$ component in (\ref{eqn:traveltime2}) is independent of the dielectric refractive index $n$. This is explained in the next section.  If the clocks were synchronised $t_{AB}$ would be known, and by changing direction of the light path, that is varying $\theta$, the magnitude of the 2nd term may be separated from the magnitude of the 1st term.   If the clocks are not synchronised, then the measured travel time ${\overline t}_{AB}=(t_B+\tau)-t_A=t_{AB}+\tau$, where $\tau$ is the unknown, but fixed,   offset between the two clocks.  But this does not change the above argument. The magnitude of $v$ and the direction of $v$ can be determined by varying $\theta$.  For a small detector the change in $\theta$ can be achieved by a  direct rotation. But for a large detector, such as De Witte's \cite{DeWitte} 1.5km RF coaxial cable experiment, the rotation was achieved by that of the earth.  The reason for using opposing propagation directions, as in Fig.\ref{fig:oneway}, and then measuring travel time differences,  is that local temperature effects cancel. This is because a common temperature change in the two adjacent cables changes the speed to the same extent, whereas absolute motion effects cause opposite signed speed changes. Then the temperature effects cancel on measuring differences in the travel times, whereas absolute motion effects are additive. Finally, after the absolute motion velocity has been determined, the two spatially separated clocks may be synchronised.
 
That the $v/c$ term in  $t_{AB}$ in  (\ref{eqn:traveltime2}) is independent of $n$ means that various techniques to do a 1st order in $v/c$ experiment  that involves using two dielectrics with different values of $n$ fail. One such experiment was by Trimmer {\it et al.}, who used a triangular interferometer, with the light path split into one beam passing through vacuum, and the other passing through glass.  No 1st order effect was seen. This is because the $v$-dependent  travel times through the glass, and corresponding vacuum distance, have the same value to 1st order in $v/c$.  On realising this Trimmer {\it et al.} subsequently withdrew their paper, see 2nd citation in \cite{Trimmer}.  Cahill\footnote{Cahill R.T.  {\it Progress in Physics}, {\bf 4}, 73-92, 2006.} performed a dual optical-fiber/RF coaxial cable experiment that was supposedly 1st order in $v/c$.  If the Fresnel drag formalism applies to both optical fibers and RF coaxial cables, then again there could not have been any $v/c$ signal in that experiment, and the observed effects must have been induced by temperature effects.  All this implies that because of the Fresnel drag effect it appears not possible to perform a $v/c$ experiment using one clock - rather two must be used, as in Fig.\ref{fig:oneway}. This, as noted above, does not require clock synchronisation, but it does require clocks that  very stable. To use one ``clock" appears then to require 2nd order in v/c detectors, but then the effect is some 1000 times smaller, and requires interferometric methods to measure the very small travel time differences, as in gas-mode and optical-fiber Michelson interferometers. It is indeed fortuitous that the early experiments by Michelson and Morley, and by Miller, were in gas mode, but not by design.  

 The Krisher optical fiber 1st order $v/c$ experiment \cite{Krisher} measured the phase differences $\phi_1$ and $\phi_2$ between the two signals travelling in different directions through very long optical fibers,  rather than the travel time variations, as the earth rotated.  This involves two phase comparators, with one at each end of the fibers. However the phases always have a multiple of $2\pi$ phase ambiguity, and in the Krisher experiment this was overlooked.  However  the timing of the maxima/minima permitted the Right Ascension (RA) of the direction of $v$ to be determined, as the direction of propagation is changed by rotation, and the result agreed with that  found by Miller; see \cite{CahillNASA} for plots of the Krisher data plotted against local sidereal times.
 
 \section{Deriving the Fresnel Drag Formalism}

 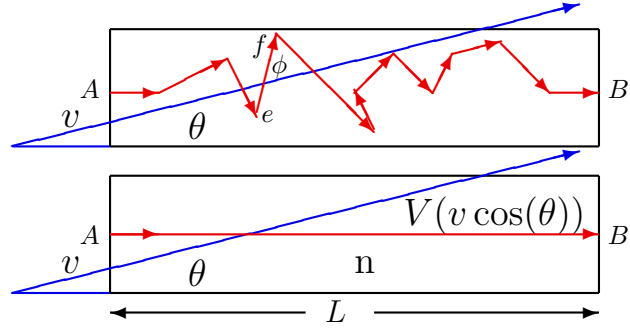
\begin{figure}
\vspace{20mm}
\hspace{0mm}
\setlength{\unitlength}{1.3mm}
\hspace{0mm}\begin{picture}(0,0)
\thicklines

\definecolor{dunkelmagenta}{rgb}{.9, 0, .0}
\definecolor{hellgrau}{gray}{.8}
\definecolor{dunkelblau}{rgb}{0, 0, .9}
\definecolor{roetlich}{rgb}{1, .7, .7}
\definecolor{dunkelmagenta}{rgb}{.9, 0, .0}
\definecolor{green}{rgb}{0, 1,0.4}
\definecolor{black}{rgb}{0, 0, 0}

\color{black}

\put(46.5,7.5){{$ \phi$}}
\put(45.5,2.5){{$e$}}
\put(44.5,9.6){{$f$}}

\put(38,0.75){{\Large $ \theta$}}

\put(38,-14.25){{\Large $ \theta$}}
\put(25,2){{\Large \bf $v$}}
\put(25,-13){{\Large \bf $v$}}

\put(60,-8){{\Large \bf $V(v\cos(\theta))$}}
\put(55,-13){{\Large n}}

\put(30,0){\line(0,1){12}}
\put(30,0){\line(1,0){50}}
\put(30,12){\line(1,0){50}}
\put(80,0){\line(0,1){12}}

\put(30,-15){\line(0,1){12}}
\put(30,-15){\line(1,0){50}}
\put(30,-3){\line(1,0){50}}
\put(80,-15){\line(0,1){12}}

\put(26,5){{ $A$}}
\put(80,5){{ $B$}}

\put(26,-10){{ $A$}}
\put(80,-10){{ $B$}}

\put(50,-17){\vector(-1,0){20}}
\put(52,-18){\large$ L$}
\put(57,-17){\vector(1,0){23}}

\color{dunkelblau}
\put(20,0){\vector(4,1){58}}
\put(20,0){\line(1,0){10}}

\put(20,-15){\vector(4,1){58}}
\put(20,-15){\line(1,0){10}}

\color{dunkelmagenta}
\put(30,-9){\vector(1,0){50}}
\put(30,-9){\vector(1,0){5}}

\color{dunkelmagenta}
\put(30,5.5){\vector(1,0){5}}
\put(35,5.5){\vector(2,1){7}}
\put(42,9){\vector(1,-2){3}}
\put(45,3.5){\vector(1,4){2}}
\put(47,11.5){\vector(1,-1){10}}
\put(57,1.8){\vector(-1,2){2}}
\put(55,5.5){\vector(1,1){4}}
\put(59,9.5){\vector(1,-1){4}}
\put(63,5.5){\vector(1,2){2}}
\put(65,9.5){\vector(4,1){5}}
\put(70,10.5){\vector(1,-1){5}}
\put(75,5.5){\vector(1,0){5}}
\end{picture}

\vspace{23mm}
\caption{\small{Slab of dielectric, length $L$,  traveling through space with velocity $\bf v$, and with EM radiation traveling, overall,  from $A$ to $B$, drawn in rest frame of slab. Top:  Microscopic model showing scattering events, with free propagation at speed $c$ relative to the space, between scattering events. Bottom: The derived macroscopic  phenomenological description showing the signal travelling at speed $V(v\cos(\theta))$, as given by the Fresnel drag expression in (\ref{eqn:Fresnel}). The dielectric refractive index is $n$.}}
 \label{fig:Fresnel}
\end{figure}

  \begin{figure*}
    \vspace{5mm}
   \hspace{0mm}\includegraphics[scale=0.295]{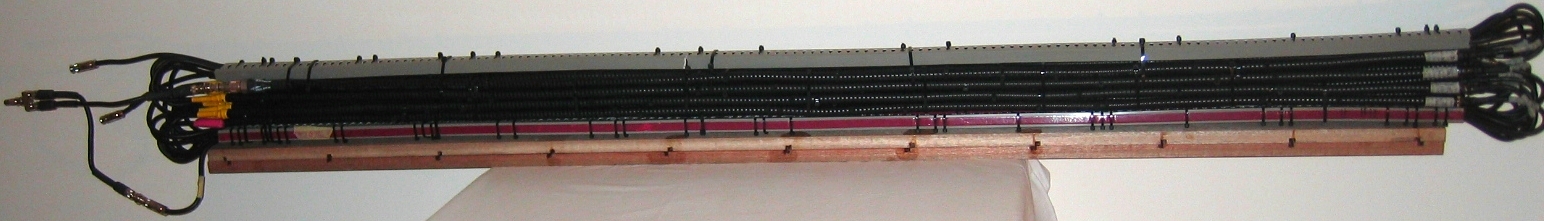}  
   \vspace{-2mm}
\caption{\small{Photograph of the RF coaxial cables arrangement, based upon 16 $\times$1.85m lengths of phase stabilised  Andrew HJ4-50 coaxial cable. These are joined to 16 lengths of FSJ1-50A cable, in the manner shown schematically in Fig.\ref{fig:DualCoax}.  The 16 HJ4-50 coaxial cables have been tightly bound into a 4$\times$4 array, so that the cables, locally, have the same temperature, with cables in one of the circuits embedded between cables in the 2nd circuit.  This arrangement of the cables permits the cancellation of temperature effects in the cables. A similar array of the smaller  diameter FSJ1-50A cables  is located inside the grey-coloured conduit  boxes.  This arrangement has permitted the study of the Fresnel drag effect in RF coaxial cables, and revealed that the usual Fresnel drag speed expression applies. }}
\label{fig:coaxphoto}\end{figure*}

Here we give a heuristic derivation of the Fresnel drag speed formalism in a moving dielectric, with the dielectric modeled by random geometrical-optics paths, see Fig.\ref{fig:Fresnel}.  These  may be thought of as modelling EM wave scatterings, and their associated time delays.
The slab of dielectric has  length $L$ and  travels through space with velocity $\bf v$, and with EM radiation traveling, overall,  from $A$ to $B$.  The top of  Fig.\ref{fig:Fresnel} shows the microscopic heuristic model of propagation through the dielectric with EM radiation traveling at speed $c$ wrt space between scattering events, being scattered from random sites - atoms, moving through space with velocity  $\bf v$.    The bottom of  Fig.\ref{fig:Fresnel} shows the macroscopic  description with EM radiation  effectively traveling in a straight line directly from $A$ to $B$,  with effective linear speed $V(v\cos(\theta))$, and with the dielectric now described by a refractive index $n$. 

The key insight is that when a dielectric has absolute velocity  ${\bf v}$ through space, the EM radiation travels at speed $c$ wrt space, between two scattering events within the dielectric. Consider a straight line propagation between scattering events $e$ and $f$, with angle $\phi$ to ${\bf v}$, see Fig.\ref{fig:Fresnel}. Consider the  paths  from the rest frame of the space. The EM wave must travel to a point in space $f^\prime$,  and then the distance travelled  $dl^\prime$, at speed $c$, is determined by the vector sum ${\bf dl^\prime}={\bf dl}+ {\bf v}dt$, with $dl$ the distance between scattering points $e$ and $f$, defined in the rest frame of the matter,  and ${\bf v} dt$ is  the displacement of $f$ to $f^\prime$, because of the absolute motion of the scattering atoms. Then the travel time to 1st order in $v/c$ is
\begin{equation}
dt =\frac{dl^\prime}{c}\approx \frac{dl}{c}+\frac{v  \cos(\phi) dt }{c}+..., \mbox{\ \ \ giving}
\end{equation}
\begin{equation}
dt \approx \frac{dl}{c}+\frac{v dl \cos(\phi)  }{c^2}+...= \frac{dl}{c}+\frac{{\bf v} \cdot  {\bf dl}   }{c^2}+...
\end{equation} 
We ignore Lorentz length contraction of the slab  as this only contributes at 2nd order in $v/c$.  
Summing over paths to get total travel time from $A$ to $B$
\begin{eqnarray}
t_{AB} &=&\int_A^B \frac{dl}{c}+\int_A^B \frac{{\bf v} \cdot  {\bf dl}   }{c^2}+... \nonumber \\
&=& \frac{l}{c} + \frac{L v \cos(\theta)}{c^2}+...\nonumber \\
&=& \frac{n L}{c} + \frac{L v \cos(\theta)}{c^2}+...,
\label{eqn:FD}\end{eqnarray} 
where $L$ is the straight line distance from $A$ to $B$ in the matter rest frame, and $n=l/L$ defines the refractive index of the dielectric in this treatment, as when the dielectric is at rest the effective speed of EM radiation through matter in a straight line from $A$ to $B$ is defined to be $c/n$. Note that $t_{AB}$   does not involve $n$ in the $v$ dependent 2nd term. This effect is actually the reason for the Fresnel drag formalism.  The macroscopic treatment, which leads to the Fresnel drag formalism,   involves the  sum $ |{\bf L}^\prime |=|{\bf  L}+{\bf v}t_{AB}|$, for the macrosocopic distance traveled, which gives for the travel time
\begin{eqnarray}
t_{AB}&=&\frac{L^\prime}{V}\approx\frac{L}{V(v\cos(\theta))} +\frac{v \cos(\theta)t_{AB}}{V(v\cos(\theta))},  \mbox{\ \ \ giving} \nonumber \\
t_{AB}&=&\displaystyle{\frac{L}{V(v\cos(\theta))} +\frac{L v \cos(\theta)}{V(v\cos(\theta))^2}}+...
\end{eqnarray}
where $V(v\cos(\theta))$ is the effective linear speed of EM radiation in direction $AB$ at angle $\theta$ to ${\bf v}$, and $v \cos(\theta)t_{AB}$ is the extra distance travelled, caused by the end $B$ moving. This  form assumes that the total distance $L^\prime$ is travelled at speed $V(v\cos(\theta))$. This reproduces the microscopic result (\ref{eqn:FD})  only if $V(v)=c/n+v(1-1/n^2)$, which is the Fresnel drag expression.  The key point is that the Fresnel drag formalism is needed to ensure, despite appearances, that the extra distance traveled due to the absolute motion of the dielectric, is travelled at speed $c$, and not at speed $c/n$, even though the propagation is within the dielectric. Hence there is no actual drag phenomenon involved, and so the nomenclature ``Fresnel drag"  is misleading.

However it was not  clear that the same analysis applied to RF coaxial cables, because of the possible effects of the conduction electrons in the inner and outer conductors.  The dual coaxial cable experiment reported herein  shows that the Fresnel drag expression also applies in this case. The Fresnel drag effect is a direct consequence of the absolute motion of the slab of matter through space, with the speed of EM radiation being $c$ wrt space itself.  A more complete derivation based on the Maxwell-Hertz equations is given in Drezet \cite{Drezet}.

   \begin{figure}[t]
   \vspace{1mm}
\hspace{10mm}\includegraphics[scale=0.4]{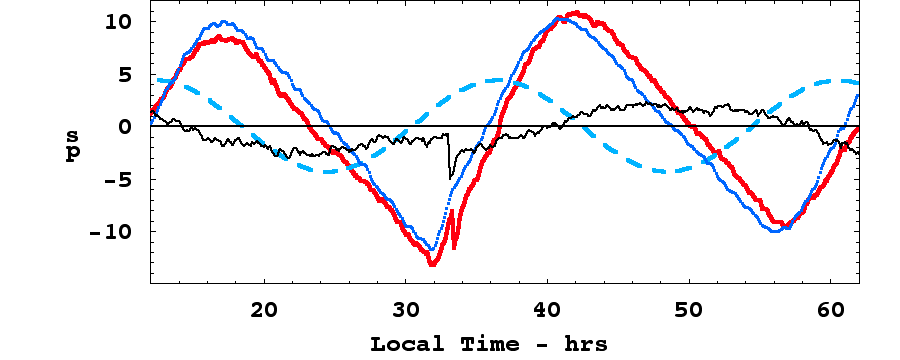} 
\setlength{\unitlength}{1.1mm}
\setlength{\unitlength}{1.1mm}
\vspace{20mm}

\hspace{40mm}\begin{picture}(0,0)
\thicklines

\definecolor{hellgrau}{gray}{.8}
\definecolor{dunkelblau}{rgb}{0, 0, .9}
\definecolor{roetlich}{rgb}{1, .7, .7}
\definecolor{dunkelmagenta}{rgb}{.9, 0, .0}
\definecolor{mauve}{rgb}{0.4, 0, .8}

\put(0,6.5){\bf S}\put(65,6.5){\bf N}

\put(5.5,-2){\bf A}
\put(67,-2){\bf B}

\put(5.5,16){\bf D}
\put(67,16){\bf C}

\put(0,0){\line(0,1){5}}\put(0,0){\line(1,0){5}}
\put(0,5){\line(1,0){5}}\put(5,0){\line(0,1){5}}
\put(0.2,1.9){\bf Rb}

\put(0,10){\line(0,1){5}}\put(0,10){\line(1,0){5}}
\put(0,15){\line(1,0){5}}\put(5,10){\line(0,1){5}}
\put(0.4,12.6){\bf DS}
\put(0.6,10.5){\bf O}

\put(20,-5){\vector(-1,0){15}}
\put(25,-5){\large$ L$}
\put(32,-5){\vector(1,0){37}}

 \color{dunkelmagenta}
  \put(32,-3){\bf FSJ1-50A}
  \put(32,8.0){\bf FSJ1-50A}
\put(5,1){\line(1,0){64}}
\put(5,1.4){\line(1,0){64}}
\put(35,1.35){\vector(1,0){1}}
 
\color{dunkelblau}
  \put(32,15){\bf HJ4-50}
    \put(32,5){\bf HJ4-50}
\put(5,4){\line(1,0){64}}
\put(5,3.6){\line(1,0){64}}

  \color{dunkelmagenta}
\put(5,11){\line(1,0){64}}
\put(5,11.4){\line(1,0){64}}

 \color{dunkelblau}
\put(5,14){\line(1,0){64}}
\put(5,13.6){\line(1,0){64}}

\put(35,13.8){\vector(-1,0){1}}
\put(35,11.2){\vector(-1,0){1}}
\put(35,3.9){\vector(1,0){1}}

 \color{dunkelmagenta}
\put(68.4,8.92){\oval(5,10.1)[rb]}
\put(68.4,6.3){\oval(5,10.1)[rt]}

\put(68.5,6.2){\oval(10,10.1)[rb]}
\put(68.5,8.8){\oval(10,10)[rt]}
\put(73.5,6){\line(0,1){4}}
\put(73.6,6){\vector(0,1){3}}
\put(70.9,6.0){\vector(0,1){2}}

\end{picture}

\vspace{5mm}\caption{\small Top: Data, from May17-19, 2010,  from the dual RF coaxial cable experiment enabling Fresnel drag in coaxial cables to be studied: red plot is relative 10 MHz RF travel times between the two circuits, and blue plot is temperature of the air  (varying from 19 to 23$^\circ$C) passing into the LeCroy DSO, scaled to fit the travel time data.  The black plot is travel time differences after correcting for DSO temperature effects. The dashed plot is time variation expected using spacecraft earth-flyby Doppler shift determined velocity, if the Fresnel drag effect is absent in RF coaxial cables. Bottom: Schematic layout of the coaxial cables. This ensures two opposing circuits that enable cancellation of local temperature effects in the cables. In practice the cables are divided further, as shown in Fig.\ref{fig:coaxphoto}. }
\label{fig:DualCoax}
\end{figure}
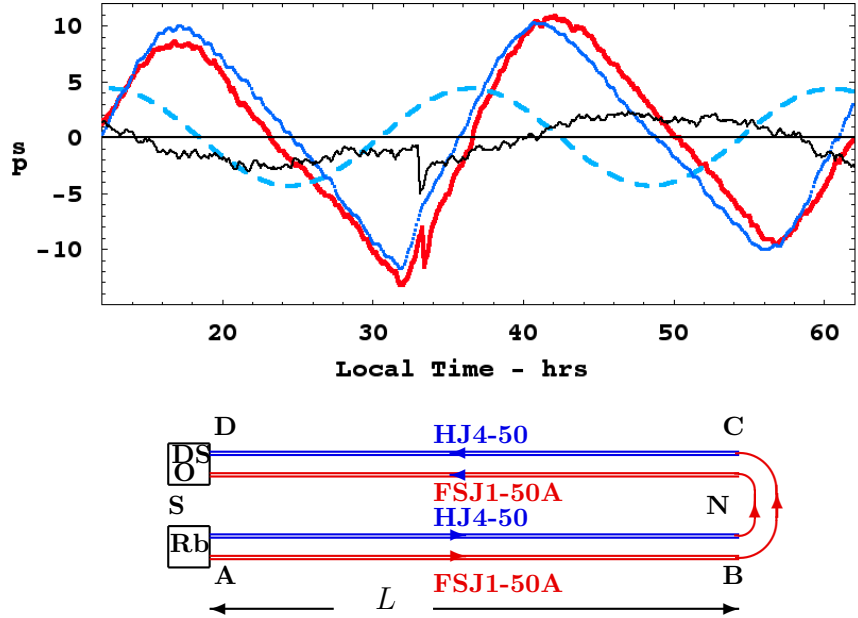

\section{Fresnel Drag Experiment in RF Coaxial Cables}
We now come to the 1st experiment that has studied the Fresnel drag effect in RF coaxial cables.  This is important for any proposed   EM anisotropy experiment using RF coaxial cables.  The query here is whether the presence of the conductors forming the coaxial cables affects the usual Fresnel drag expression in (\ref{eqn:Fresnel}), for a coaxial cable has an inner and outer conductor, with a dielectric in between.

Fig.\ref{fig:DualCoax} shows the schematic  arrangement using  two different RF coaxial cables, with two separate circuits, and Fig.\ref{fig:coaxphoto} a photograph.  One measures the travel time difference of  two RF signals from a Rubidium frequency standard (Rb) with a Digital Storage Oscilloscope (DSO). In each circuit the RF signal travels one-way in one type of coaxial cable, and returns via a different kind of coaxial cable.  Two circuits are used so that temperature effects cancel - if a temperature change alters the speed in one type of cable, and so the travel time, that travel time change is the same in both circuits, and cancels in the difference.    Even though phase-stabilised coaxial cables are used, the temperature effects  need to be cancelled in order to be able to reliably measure time differences at ps levels.  To ensure  cancellation of temperature effects, and also for practical convenience, the  Andrew HJ4-50 cables are cut into 8 $\times$ 1.85m shorter lengths in each circuit, corresponding to a net length of $L=8\times 1.85 =14.8$m. The curved parts of the Andrew FSJ1-50A cables contribute only at 2nd order in $v/c$.   

To analyse the experimental data  we modify the Fresnel drag speed expression in  (\ref{eqn:Fresnel}) to
\begin{equation}
V(v)=\frac{c}{n_i}+v\left(1-\frac{1}{m_i^2}\right)
\label{eqn:ModFresnel}\end{equation}
for each cable, $i=1,2$, where $m_i=n_i$ gives the normal Fresnel drag, while $m_i=1$ corresponds to no Fresnel drag. Repeating the derivation leading to (\ref{eqn:traveltime2}) we obtain to 1st order in $v/c$ the travel time difference between the two circuits,
\begin{equation}
\Delta t = \frac{2Lv\cos(\theta)}{c^2}\left(\frac{n_1^2}{m_1^2} -\frac{n_2^2}{m_2^2}  \right)
\label{eqn:dualtime}\end{equation}
The apparatus was orientated NS and used the rotation of the earth to change the angle $\theta$.  Then $\theta$ varies between $\lambda+\delta-90^\circ=20^\circ$ and $\lambda-\delta+90^\circ=50^\circ$, where $\lambda=35^\circ$ is the latitude of Adelaide, and $\delta=75^\circ$ is the declination of the 3-space flow from the flyby Doppler shift analysis, and with a speed of 486km/s.. Then if $m_i\neq n_i$ a signal with period 24hrs should be revealed. We need to compute the magnitude of the time difference signal if there is no Fresnel drag effect.  The FSJ1-50A has an effective refractive index $n_1=1.19$, while the HJ4-50 has $n_2=1.11$, and then $\Delta t$ would change by 8.7ps over 24 hours, and have the phase shown in Fig.\ref{fig:DualCoax}.  However while cable temperature effects have been removed by the cable layout, another source of temperature effects is from the LeCroy WaveRunner WR6051A DSO. To achieve ps timing accuracy and stability the DSO was operated in Random Interleaved Sampling (RIS) mode. This uses many signal samples to achieve higher precision.  However in this mode the DSO temperature compensation re-calibration facility is disabled. To correct for this it was discovered that the timing errors between the two DSO channels very accurately tracked the temperature of the cooling air being drawn into the DSO. Hence during the experiment that air temperature was recorded.  The Rb frequency standard was a Stanford  Research Systems FS725. t The results for  48 hours in mid May, 2010,  are shown in Fig.\ref{fig:DualCoax}: the red plot, with glitch, shows the DSO measured time difference values, while the blue plot shows the temperature variation of the DSO air-intake temperature, scaled to the time data. We see that the time data very closely tracks the air-intake temperature. Subtracting this temperature effect we obtain the smaller plot, which has a range of  5ps, but showing no 24h period. The corrected timing data may still have some small temperature effects. The glitch in the timing data near local time of 34hs was probably caused by a mechanical stress-release event in the cables.   Hence the data implies that there is no 1st order effect in v/c, and so, from (\ref{eqn:dualtime}), that $n_1/m_1=n_2/m_2$,  with the simplest interpretation being that, in each cable $m=n$. This means that the Fresnel drag effect expression in (\ref{eqn:Fresnel}) applies to RF coaxial cables.

\section{Conclusions}

The first experiment to study the Fresnel drag effect in RF coaxial cables has  revealed that  these cables exhibit the same effect as seen in bulk dielectrics and in optical fibers, and so this effect is very general, and in the case of the RF coaxial cables, is not affected by the conductors integral to RF  coaxial cables.  Because this experiment is a null experiment,  after correcting for temperature effects in the DSO, its implications follow only when the results are compared with non-null experiments. Here we have compared the results with those from the spacecraft earth-fly Doppler shift data results.  Then we can deduce that the null result is caused by the Fresnel drag effect in the cables, and not by the absence of light speed anisotropy. This is be understood from the heuristic derivation given herein, where it was shown that the Fresnel drag expression actually involves no drag effect at all, rather its form is such as to ensure that between scatterings the EM waves travel at speed $c$ wrt to the 3-space, that is, that  the ``speed of light"  is not an invariant.  This experiment, as have many others, shows that the speed of light, as measured by an observer, actually depends on the speed of that observer wrt to  3-space.   We have also shown how the speed of light may be measured in a one-way 1st order in $v/c$ experiment,  using spatially separated clocks that are not {\it a priori} synchronised, by rotating the apparatus.  Subsequently, once the velocity of space past the detector is known, the clocks may be synchronised.

Special  thanks to Professor Igor Bray for supporting this research.

\end{document}